\documentclass[twocolumn,showpacs,preprintnumbers,amsmath,amssymb]{revtex4}

\usepackage{graphicx}
\usepackage{dcolumn}
\usepackage{bm}
\def\jpsi{{J/\psi}}
\def\mop{{\langle\mathcal{O}^h_n\rangle}}
\def\OP#1#2#3#4{{\bigl.^{#1}\hspace{-1mm}{#2}_{#3}^{[#4]}}}
\def\ME#1#2#3{{\langle\mathcal{O}^{#1}_{#2}#3\rangle}}

\def\be{\begin{equation}}
\def\ee{\end{equation}}
\def\bea{\begin{eqnarray}}
\def\eea{\end{eqnarray}}
\def\NO{\nonumber}
\def\gev{\mathrm{~GeV}}

\def\dfrac{\displaystyle\frac}

\def\md{\mathrm{d}}

\def\co{{\cal O}}
\def\a{\alpha}

\def\s{\sigma}

\begin{document}


\title{Polarization for Prompt $J/\psi$ and $\psi(2s)$ Production at the Tevatron and LHC}


\author{Bin Gong$^{1,2,3}$, Lu-Ping Wan$^{1,2}$, Jian-Xiong Wang$^{1,2}$ and Hong-Fei Zhang$^{1,2}$}%
\affiliation{
Institute of High Energy Physics, Chinese Academy of Sciences, P.O. Box 918(4), Beijing, 100049, China.\\
Theoretical Physics Center for Science Facilities, CAS, Beijing, 100049, China.\\
Institute of Theoretical Physics, Chinese Academy of Sciences, P.O. Box 2735, Beijing, 100190, China.
}%

\begin{abstract}
With nonrelativistic QCD factorization, we present the first complete 
next-to-leading order study on the polarization of prompt $J/\psi$ hadroproduction 
by including feeddown from $\chi_c(^3P_J^1,^3S_1^8)$ 
and $\psi(2s)$ which turn out to be 
very important parts.  By using the color-octet long-distance matrix elements obtained from a 
combined fit of the measurements at the Tevatron and LHC 
for $J/\psi$, $\psi(2s)$ and $\chi_c$, 
the prompt $J/\psi$ polarization predictions are presented, and the results are in agreement with the CDF Run I data 
(except two points), but in conflict with the CDF run II data, while they are close to the ALICE data (inclusive $J/\psi$).   
The measurements at the LHC are expected to clarify the situation.  
\end{abstract}

\maketitle

Quantum chromodynamics (QCD) is a successful theory to describe strong interaction,
but its fundamental ingredients, the quarks and gluons, are not observed freely and must hadronize eventually.
The fact makes it impossible to calculate any processes involving detected hadrons in the final or initial states directly.
Therefore a suitable factorization scheme to divide problems into perturbative calculable and nonperturbative parts is very important.
In order to remedy the large discrepancy between theoretical prediction and experimental data on the transverse momentum ($p_t$) distribution of $\jpsi$ production at the Tevatron, 
a color-octet (CO) mechanism~\cite{Braaten:1994vv} was proposed based on nonrelativistic QCD (NRQCD) factorization~\cite{Bodwin:1994jh},
which is proven in next-to-next-to-leading order analysis~\cite{Nayak:2005rt}.
It allows consistent theoretical prediction to be made and improved perturbatively. 
However, the leading-order calculation based on NRQCD predicts a sizable transverse polarization for $\jpsi$ hadroproduction at high $p_t$~\cite{Beneke:1995yb}
while the measurement at the Tevatron~\cite{Abulencia:2007us} gives almost an unpolarized result.
Many efforts have been made to clarify the situation thereafter (see Ref.~\cite{ConesadelValle:2011fw} and references therein).
It is expected that the long-standing $\jpsi$ polarization puzzle could be solved or clarified by the experimental measurements at the LHC and progress in theoretical calculation. 

In the last five years, there was some very important progress made in the next-to-leading order (NLO) QCD correction calculation.  
The NLO corrections to color-singlet $\jpsi$ hadroproduction have been investigated in Refs.~\cite{Campbell:2007ws,Gong:2008sn}
and its $p_t$ distribution is found to be enhanced by $2-3$ order of magnitude at the high $p_t$ region.
And it is found that $\jpsi$ polarization changes from transverse into longitudinal at NLO~\cite{Gong:2008sn}. 
The results are reproduced at leading-order in a new factorization scheme for large $p_t$ quarkonium production~\cite{Kang:2011mg}.
The NLO corrections to $\jpsi$ production via $S$-wave CO states ($\OP{1}{S}{0}{8}, \OP{3}{S}{1}{8}$) are studied in Ref.~\cite{Gong:2008ft}
and the corrections to $p_t$ distributions of both $\jpsi$ yield and polarization are small.
In Refs.~\cite{Ma:2010vd}, NLO corrections for $\chi_{cJ}$ hadroproduction are studied.
The complete NLO calculation for prompt $\jpsi$ hadroproduction (with $\OP{3}{P}{J}{8}$ included) is given by two groups~\cite{Ma:2010yw, Butenschoen:2010rq},
and their predictions for $p_t$ distributions agree with the experimental measurements at the Tevatron and LHC.
Similar progress is also achieved for $\Upsilon$ hadroproduction\cite{Artoisenet:2008fc} and $\jpsi$ photoproduction~\cite{Artoisenet:2009xh}.

Recently, the complete NLO calculation for polarization of direct $\jpsi$ ($\OP{3}{P}{J}{8}$ included) hadroproduction are presented by two groups~\cite{Butenschoen:2012px, Chao:2012iv}.
One of them~\cite{Butenschoen:2012px} find that the CDF data (run II)~\cite{Abulencia:2007us} cannot be interpreted by using a global fit of $\jpsi$ yield data,
while the other~\cite{Chao:2012iv} finds that it can be interpreted by using a combined fit of $\jpsi$ yield and polarization data from the CDF.
However, there only exist polarization measurements for prompt (or even inclusive) $\jpsi$ production until now~\cite{Affolder:2000nn,Abulencia:2007us}.
It is known that among all the feeddown contributions to prompt $\jpsi$ production from higher charmonium states,
$\chi_{cJ}$ contributes more than $20-30\%$ of prompt experimental measurements~\cite{Abe:1997yz, LHCb:2012af},
and $\psi(2s)$ also contributes a small fraction, while others are negligible.
The feeddown contribution is so large that it can drastically change the polarization results and must be considered.  
Therefore, to test NRQCD factorization and solve (or clarify) the long-standing $\jpsi$ polarization puzzle,
it is a very important step to achieve the polarization predictions for prompt $\jpsi$ hadroproduction. 

In this Letter, we present the calculation of the polarization for prompt $\jpsi$ hadroproduction at NLO QCD. 
In addition to calculating the polarization of $\jpsi$ from the same intermediate states ($\OP{3}{S}{1}{1}$, $\OP{1}{S}{0}{8}$, $\OP{3}{S}{1}{8}$, 
$\OP{3}{P}{J}{8}$) as in previous works~\cite{Gong:2008sn, Gong:2008ft, Butenschoen:2012px, Chao:2012iv}, 
we have also calculated that from $\chi_{cJ}$ ($\OP{3}{P}{J}{1}$, $\OP{3}{S}{1}{8}$) feeddown,
in which the calculation of the $\OP{3}{P}{J}{1}$ part is even more complicated than that of the $\OP{3}{P}{J}{8}$ part of $\jpsi$. 
In the presented results, the polarization for $\psi(2s)$,$\chi_{cJ}$ and prompt $\jpsi$ are obtained for the first time.

\label{calculation}
In NRQCD formalism, the cross section of $h$ hadroproduction is 
\bea
\s[pp\rightarrow hx]=\sum\int dx_1 dx_2 G^i_pG^j_p \hat{\s}[ij\rightarrow (c\bar{c})_nx]\mop,
\eea \
where $p$ is either a proton or antiproton,
the indices $i, j$ run over all the partonic species and $n$ represents the $c\bar{c}$ intermediate states
($\OP{3}{S}{1}{1}$, $\OP{3}{S}{1}{8}$, $\OP{1}{S}{0}{8}$, $\OP{3}{P}{J}{8}$) for $\jpsi$ and $\psi(2s)$,
or ($\OP{3}{P}{J}{1}$, $\OP{3}{S}{1}{8}$) for $\chi_{cJ}$. 
The short-distance contribution  $\hat{\s}$ can be perturbatively calculated and the long-distance matrix elements (LDMEs) $\mop$ are governed by nonperturbative QCD effects. 

The polarization of $\jpsi$ is described in Ref.~\cite{Beneke:1998re} as
\be
\lambda=\frac{\md\sigma_{11}-\md\sigma_{00}}{\md\sigma_{11}+\md\sigma_{00}},
\mu=\frac{\sqrt{2}\mathrm{Re}\md\sigma_{10}}{\md\sigma_{11}+\md\sigma_{00}},
\nu=\frac{2\md\sigma_{1,-1}}{\md\sigma_{11}+\md\sigma_{00}}.\NO
\ee
where $\md\sigma_{S_zS_z^\prime}$  is the spin density matrix of $\jpsi$ hadroproduction.
For experimental measurement,
a simultaneous study of the polarization variables in different reference frames is particularly interesting since consistency checks on the results can be performed,
and it provides a much better control of the systematic effects due to detector limitations and analysis biases. 
Therefore, both the helicity and Collins-Soper (CS) frames are employed in our polarization calculation. 
To obtain $\md\sigma_{S_zS_z^\prime}$, 
similar treatment as in Ref.~\cite{Artoisenet:2009xh} is taken for direct $\jpsi$ and $\psi(2s)$ production. 
The feeddown from $\psi(2s)$ is obtained from $\md\sigma^{\jpsi}_{S_zS_z^\prime}|_{\psi(2s)}=\md\sigma^{\psi(2s)}_{S_zS_z^\prime} {\cal B}(\psi(2s)\rightarrow\jpsi)$.
$\chi_{c0}$, $\chi_{c1}$, and  $\chi_{c2}$ have different masses and decay branching ratios to $\jpsi$, thus are treated differently from $\jpsi(\OP{3}{P}{J}{8})$. 
The feeddown can be expressed as
\be
\langle J/\psi(S_z)\gamma(L_z)|\chi_{c}(J,J_{z})\rangle\equiv a_{J}C^{S_z,L_z}_{J,J_{z}}.
\ee
Here, $C^{S_z,L_z}_{J,J_{z}}$ is the Clebsch-Gordan coefficient, and $a_J$ is supposed to be independent of $S_z$, $L_z$, and $J_z$ as an approximation.
Then the branching ratio of $\chi_{cJ}$ to $\jpsi$ is approximately expressed as
\bea
{\cal B}(\chi_{cJ}\rightarrow\jpsi)
=\sum_{J_z,L_z,S_z}\left|a_{J}C^{S_z,L_z}_{J,J_{z}}\right|^2=|a_{J}|^2,
\eea
from which we obtain $|a_{J}|^2$.
Thus, the spin density matrix of $J/\psi$ from $\chi_{cJ}$ feeddown is obtained as
\bea
\md\sigma^{\jpsi}_{S_zS_z^\prime}|_{\chi_{cJ}} &=&{\cal B}(\chi_{cJ}\rightarrow\jpsi) \sum_{J_z,J_z^\prime}
\delta_{J_z-S_z,J_z^\prime-S_z^\prime} \nonumber\\&\times&C^{S_z,J_z-S_z}_{J,J_z} C^{*S_z^\prime,J_z^\prime-S_z^\prime}_{J,J_z^\prime} \md\sigma^{\chi_{cJ}}_{J_z,J_z^\prime}.
\eea

The newly upgraded FDC package~\cite{Wang:2004du} is used in our calculation, 
in which the reduction method for loop integrals 
proposed in Ref.~\cite{Duplancic:2003tv} is implemented. 

\label{numerical results}
In our numerical calculation, The parton distribution function CTEQ6M~\cite{Pumplin:2002vw} and the corresponding two-loop QCD coupling constant $\a_s$ are used. 
The charm-quark mass is chosen as $m_c=1.5\gev$ and an approximation $M_h=2m_c$ is made to fix the masses of quarkonia. 
The color-singlet LDMEs are estimated by using a potential model result~\cite{Eichten:1995ch},
which gives $|R_{J/\psi}(0)|^2=0.810 \gev^3$, $|R_{\psi(2s)}(0)|^2=0.529\gev^3$, and $|R'_{\chi_c}(0)|^2=0.075\gev^5$, respectively.
Branching ratios are ${\cal B}(J/\psi[\psi(2s)]\rightarrow\mu\mu)=0.0593(0.0077)$,
${\cal B}[\psi(2s)\rightarrow J/\psi]=0.595$ and ${\cal B}(\chi_{cJ}\rightarrow J/\psi)=0.0116,0.344,0.195$ for $J=0,1,2$, respectively~\cite{Nakamura:2010zzi}. 
The factorization, renormalization and NRQCD scales are chosen as $\mu_{r}=\mu_{f}=\sqrt{4m_{c}^{2}+p_t^{2}}$ and $\mu_{\Lambda}=m_c$,
respectively.  The center-of-mass energies are 1.96 and 7 TeV for the Tevatron and LHC, respectively.
It is well known that the uncertainties for $p_t$ distribution of charmonium hadroproduction from $m_c, \mu_{\Lambda}, \mu_r$ and $\mu_f$ are large at small $p_t$ region.
A recent work on relativistic corrections to $\jpsi$ hadroproduction~\cite{Xu:2012am} also shows that the correction is negative and large when $p_t<10\gev$. 
It is also very clearly shown in Refs.~\cite{Gong:2008ft,Ma:2010yw} that the experimental data in small $p_t$ region can not be interpreted well.
In Ref.~\cite{Butenschoen:2010rq}, the data in small $p_t$ region are also included in their fit, but the experimental data in large $p_t$ region are sacrificed.
Therefore, data in $p_t<7\gev$ region are excluded in our fit.

\label{fit}
\begin{figure*}
\center{
\includegraphics*[scale=0.31]{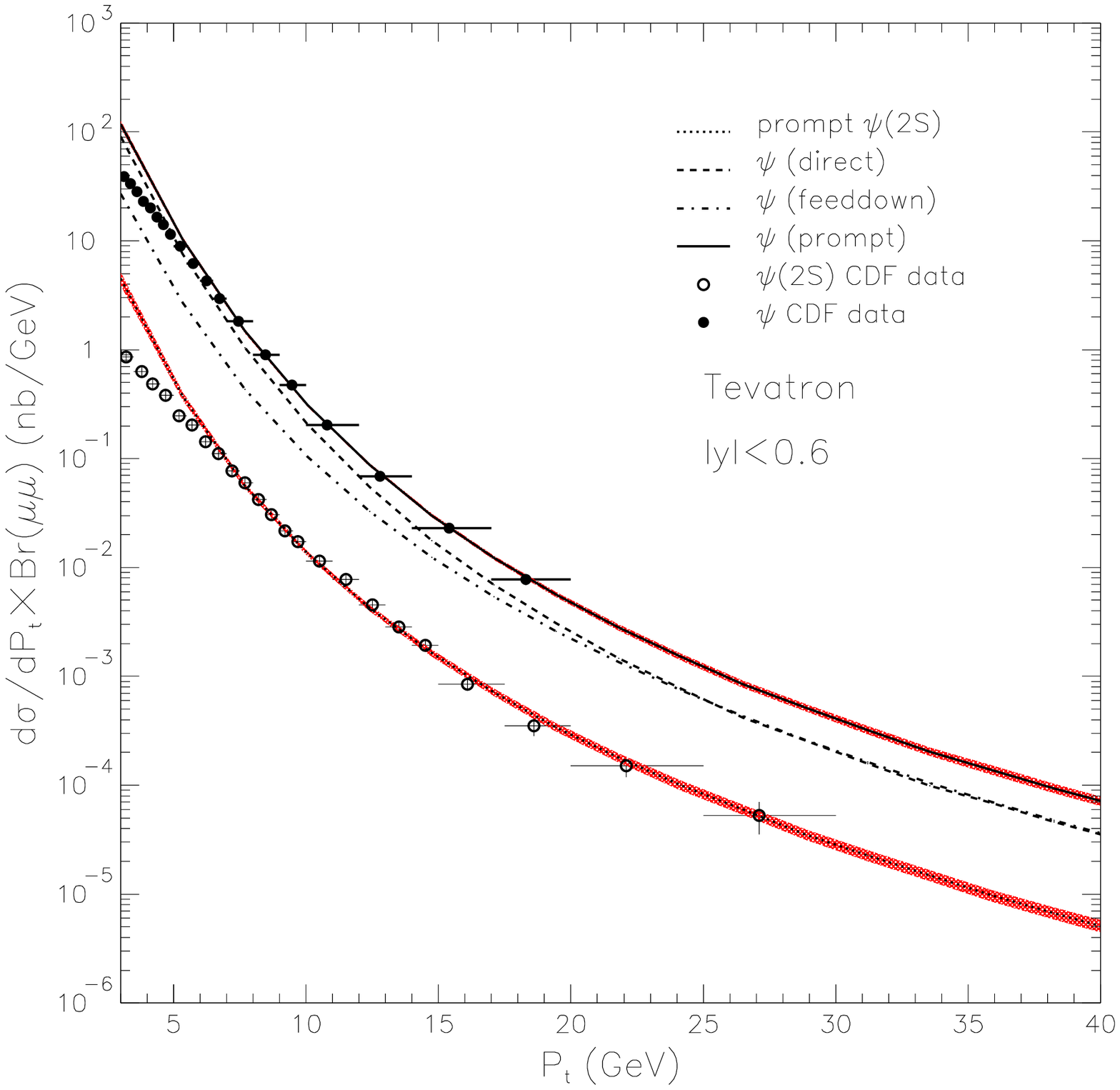}
\includegraphics*[scale=0.31]{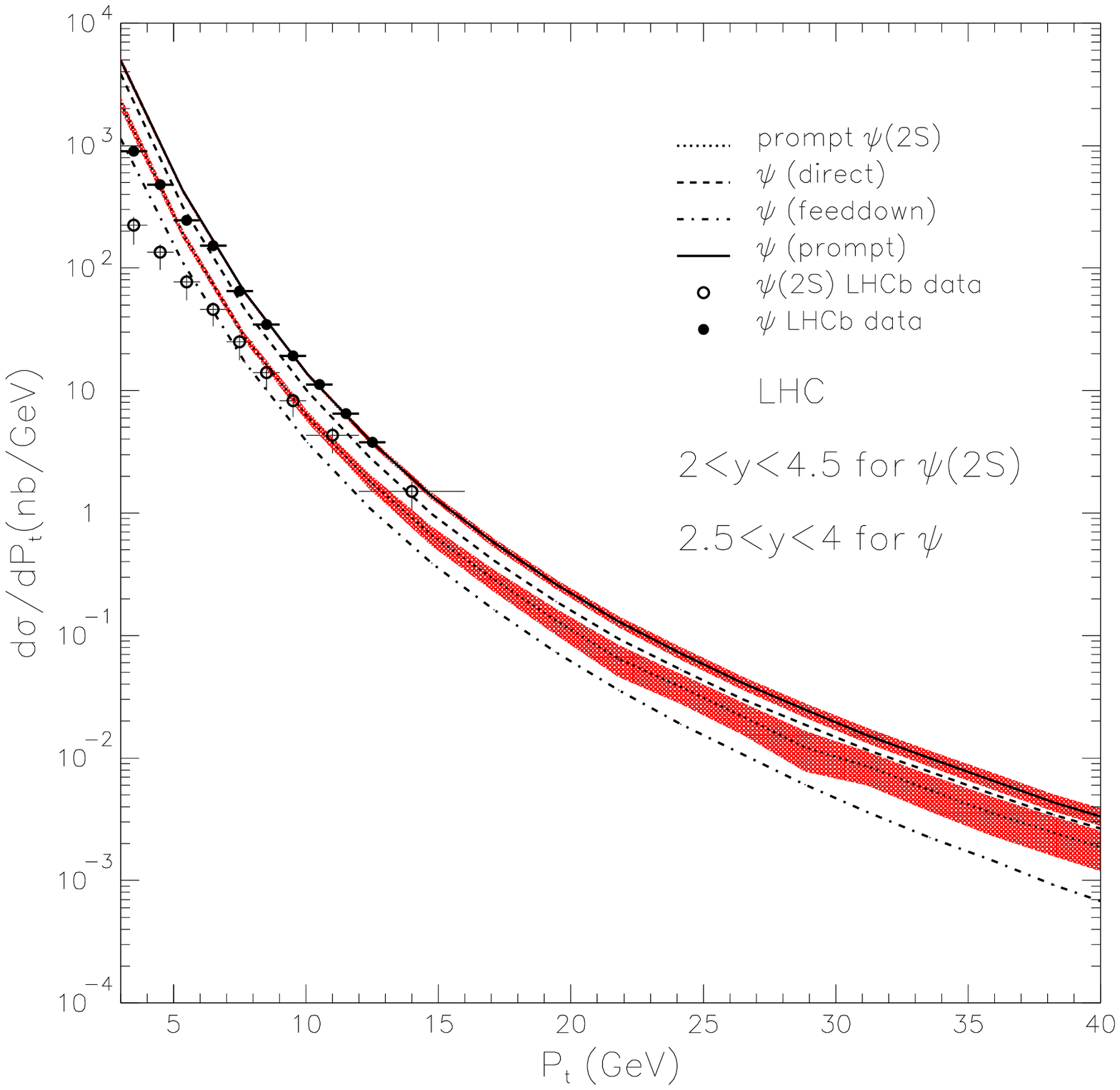}
\caption {\label{fig:fit}$p_t$ distribution of prompt $\jpsi$ and $\psi(2s)$ hadroproduction. CDF and LHCb data are taken from Refs.~\cite{Acosta:2004yw, Aaltonen:2009dm, Aaij:2011jh}.
}}
\end{figure*}

In Ref.~\cite{Chao:2012iv}, a combined fit of the prompt $\jpsi$ yield and polarization measurements of CDF (run II) is performed.
Without the feeddown contribution, we can repeat their results by an exactly same fit. 
With the feeddown contribution,  we perform a similar fit and find that the $p_t$ distribution of transverse polarized production rate for direct $\jpsi$ part becomes negative when $p_t>24\gev$, 
by varying the values of fitted CO LDMEs within the uncertainty range, the negative behavior of direct part can be delayed as $p_t$ increases, 
but cannot be avoided, which means that no physical solution in the LDME parameter space can be achieved.
It clearly shows that the feeddown contribution is so important that the conclusions for the polarization are completely different for the cases with or without the feeddown part.
Therefore, we choose to fit the CDF and LHCb experimental data for the yield only and give predictions for polarization in this work.
In addition, $p_t^{\jpsi}\approx p_t^h(M_{\jpsi}/M_h)$ is used as an approximation to count for the kinematics effect in the feeddown of $h=\psi(2s),\chi_{cJ}$.

\begin{figure*}
\center{
\includegraphics*[scale=0.31]{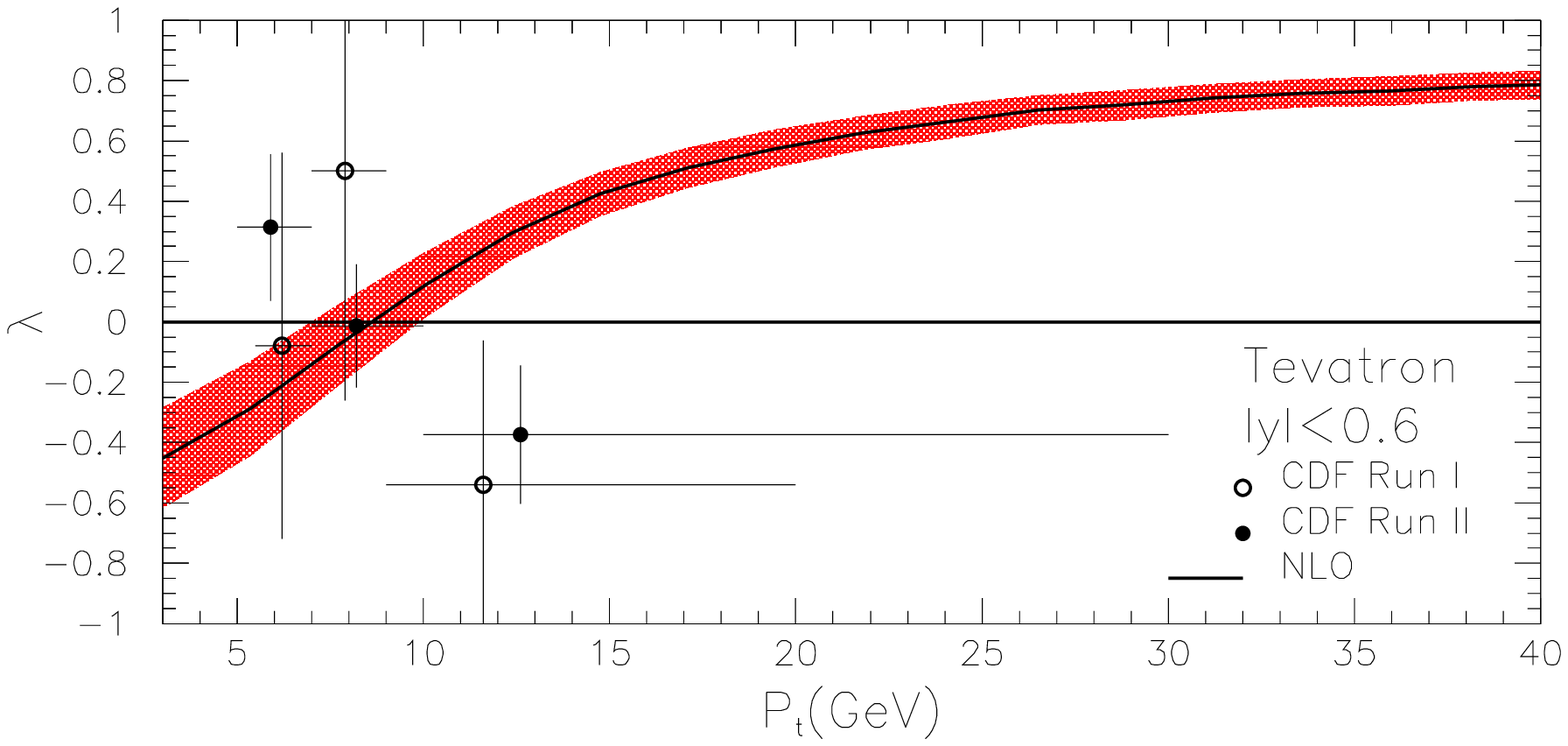}
\includegraphics*[scale=0.31]{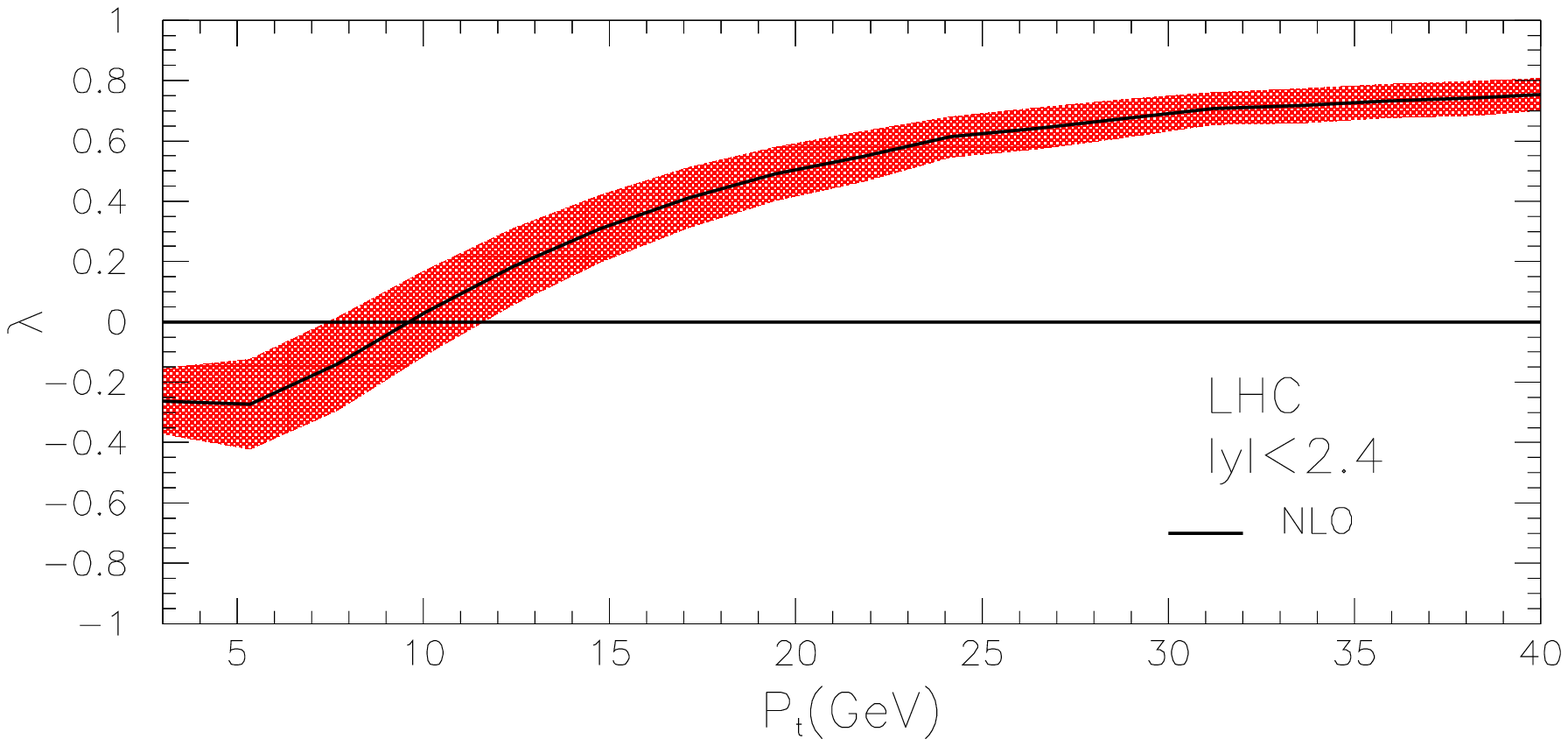}
\includegraphics*[scale=0.31]{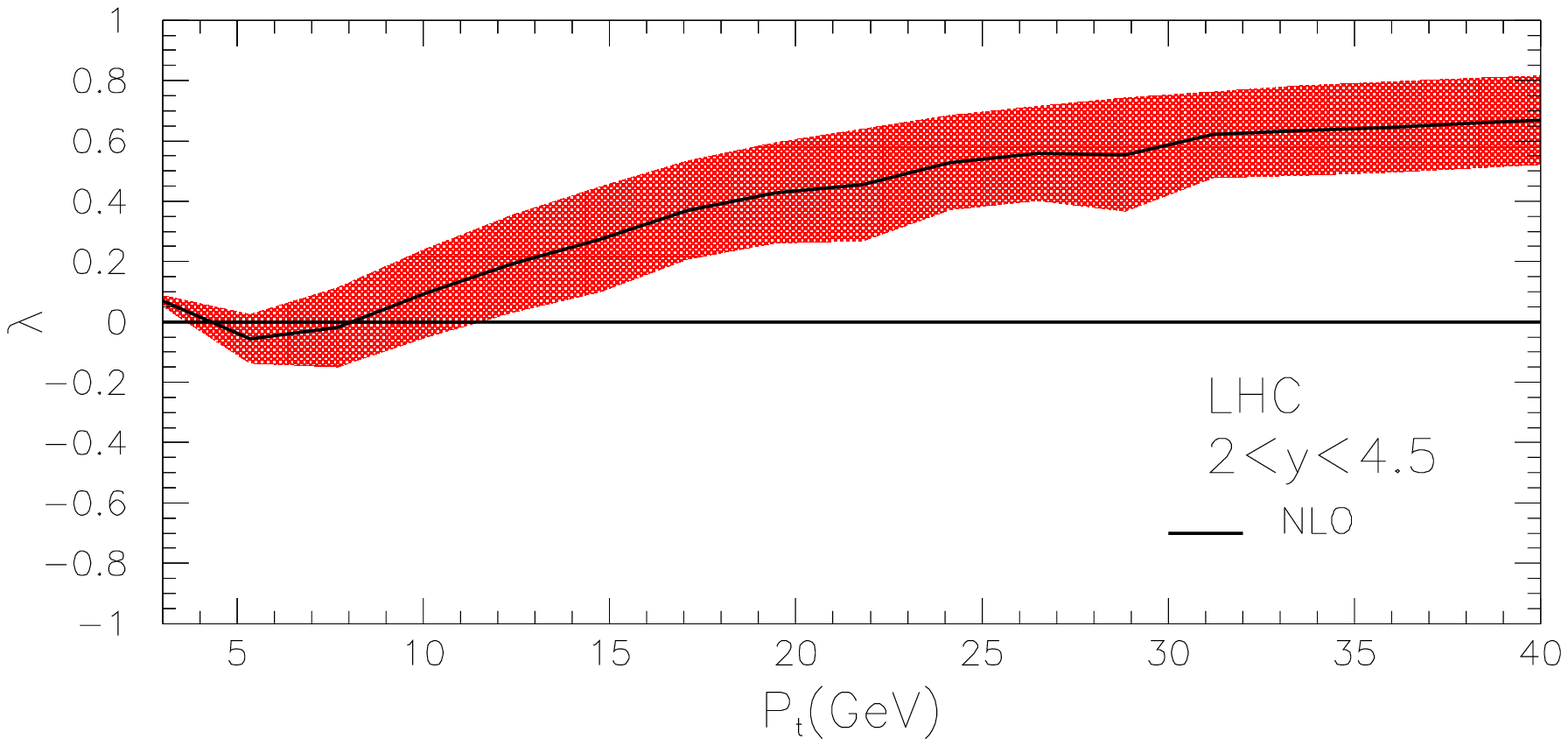}
\\
\includegraphics*[scale=0.31]{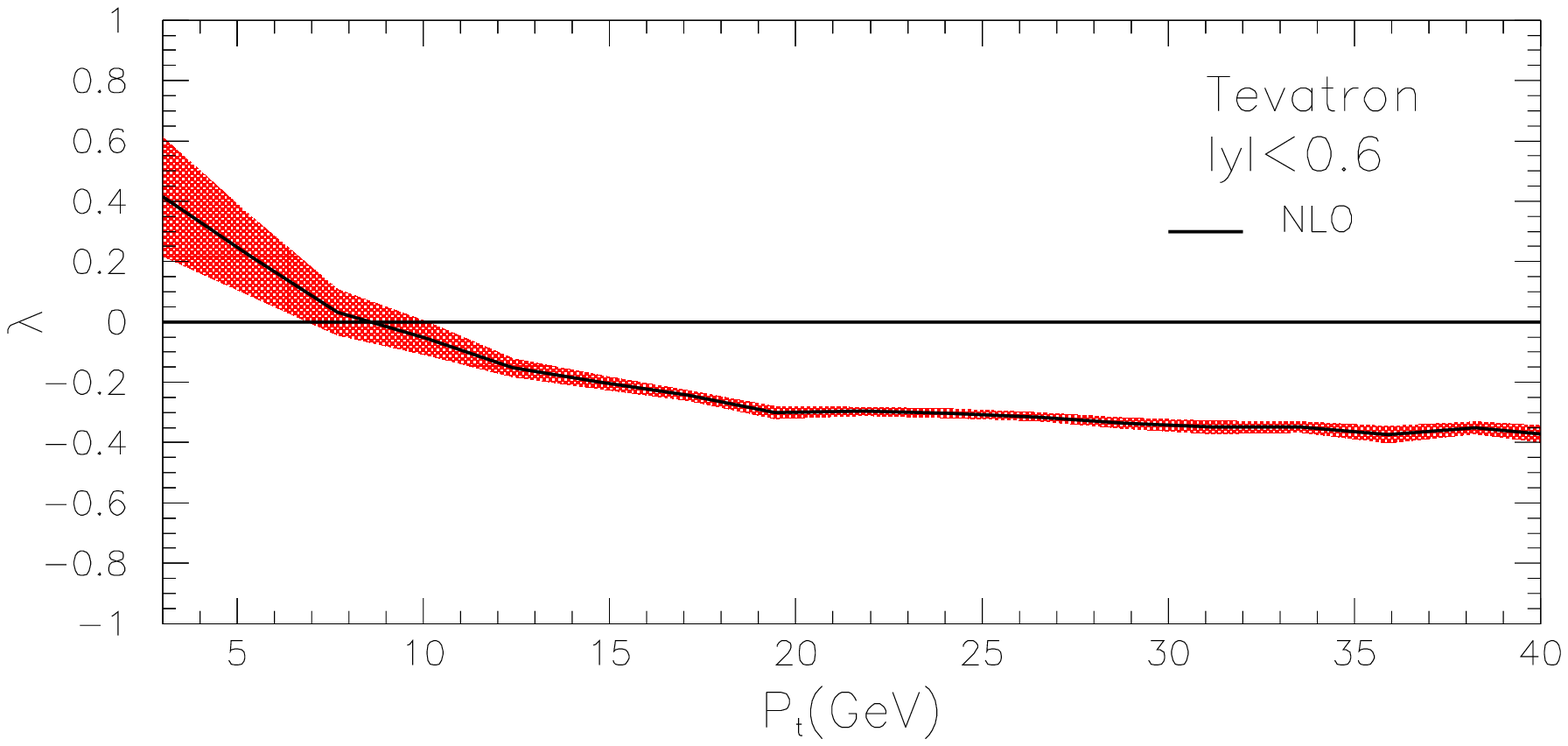}
\includegraphics*[scale=0.31]{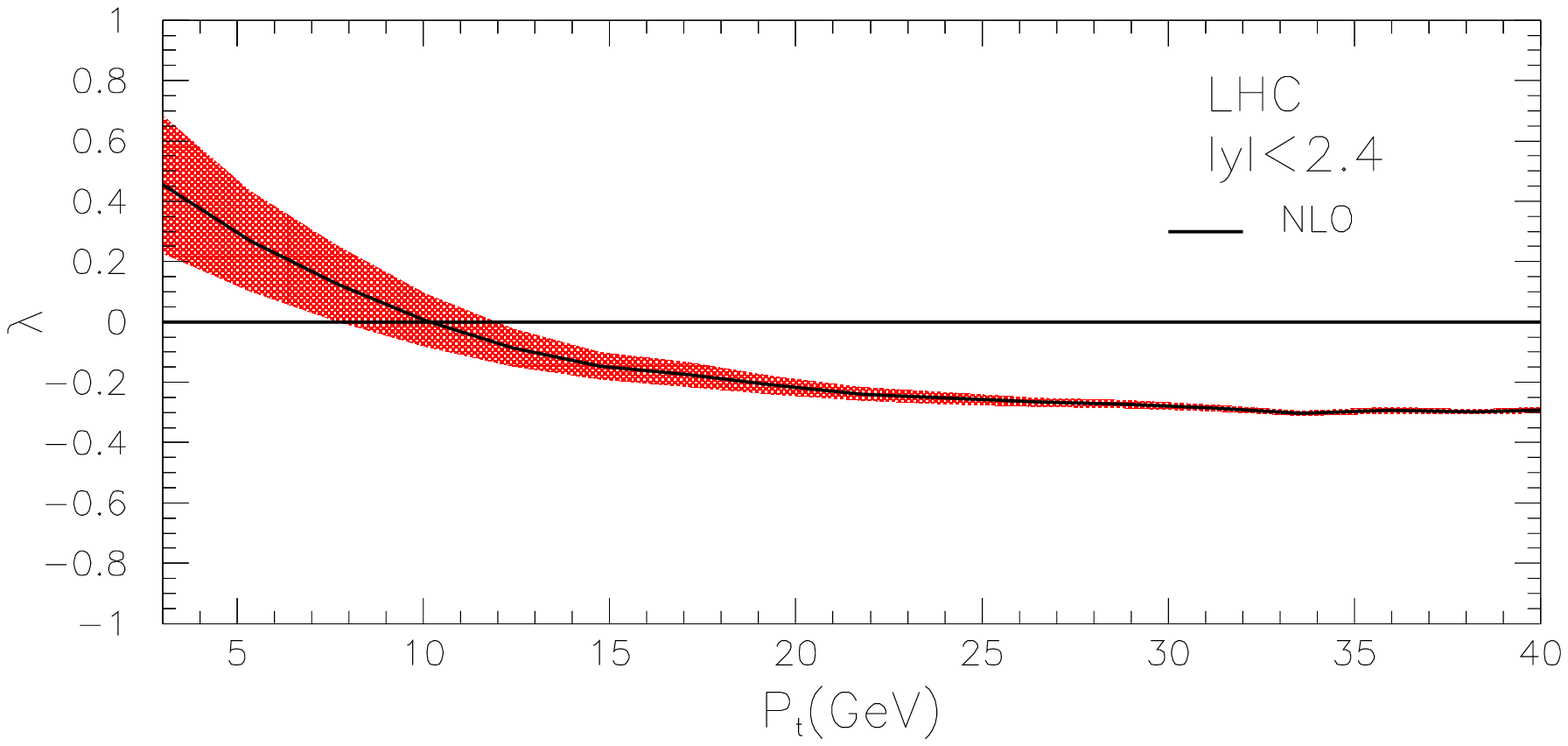}
\includegraphics*[scale=0.31]{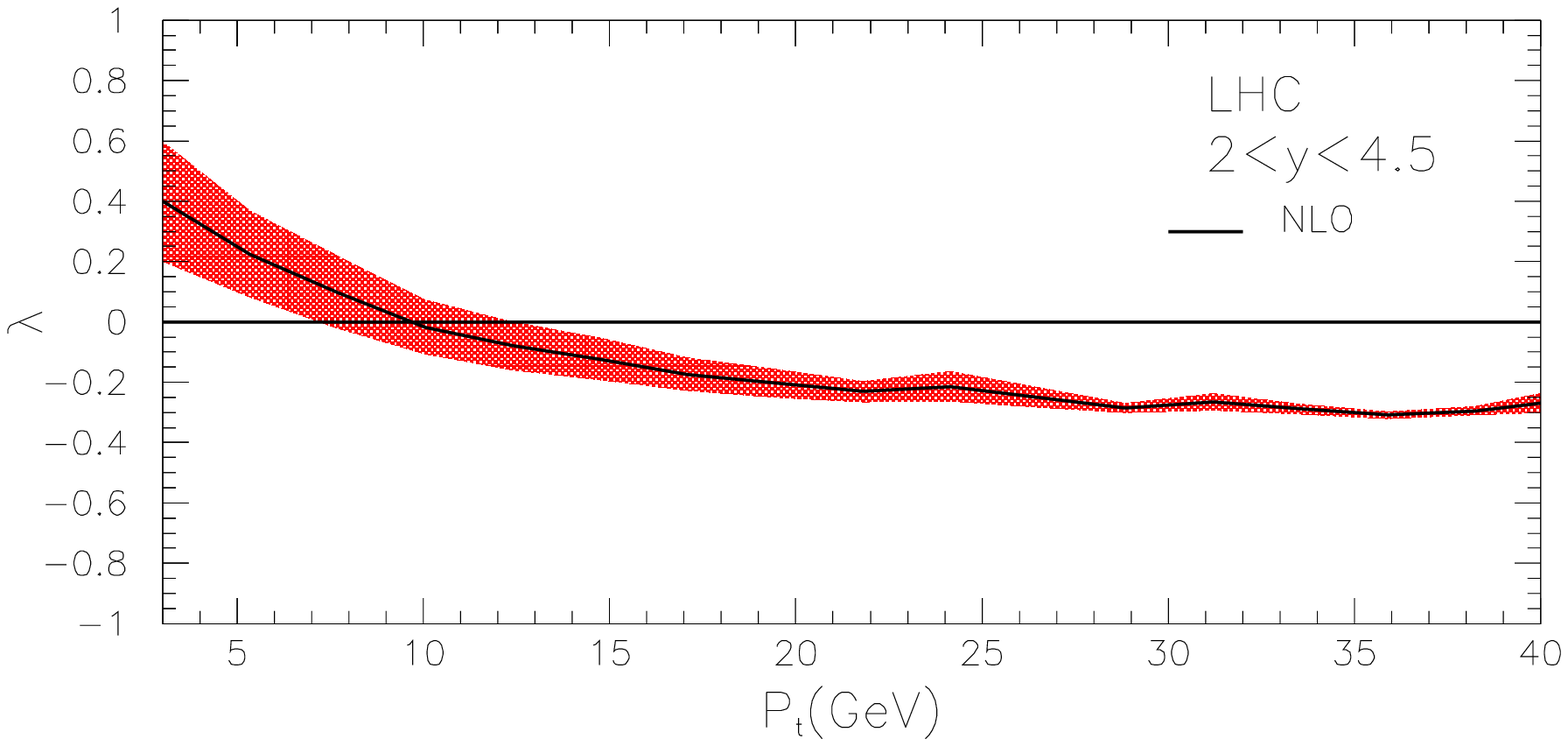}
\caption {\label{fig:pol2}Polarization parameter $\lambda$ of $\psi(2s)$ in helicity(up) and CS(down) frames. The CDF data are taken from Ref.~\cite{Abulencia:2007us,Affolder:2000nn}
}}
\end{figure*}

\begin{figure*}
\center{
\includegraphics*[scale=0.31]{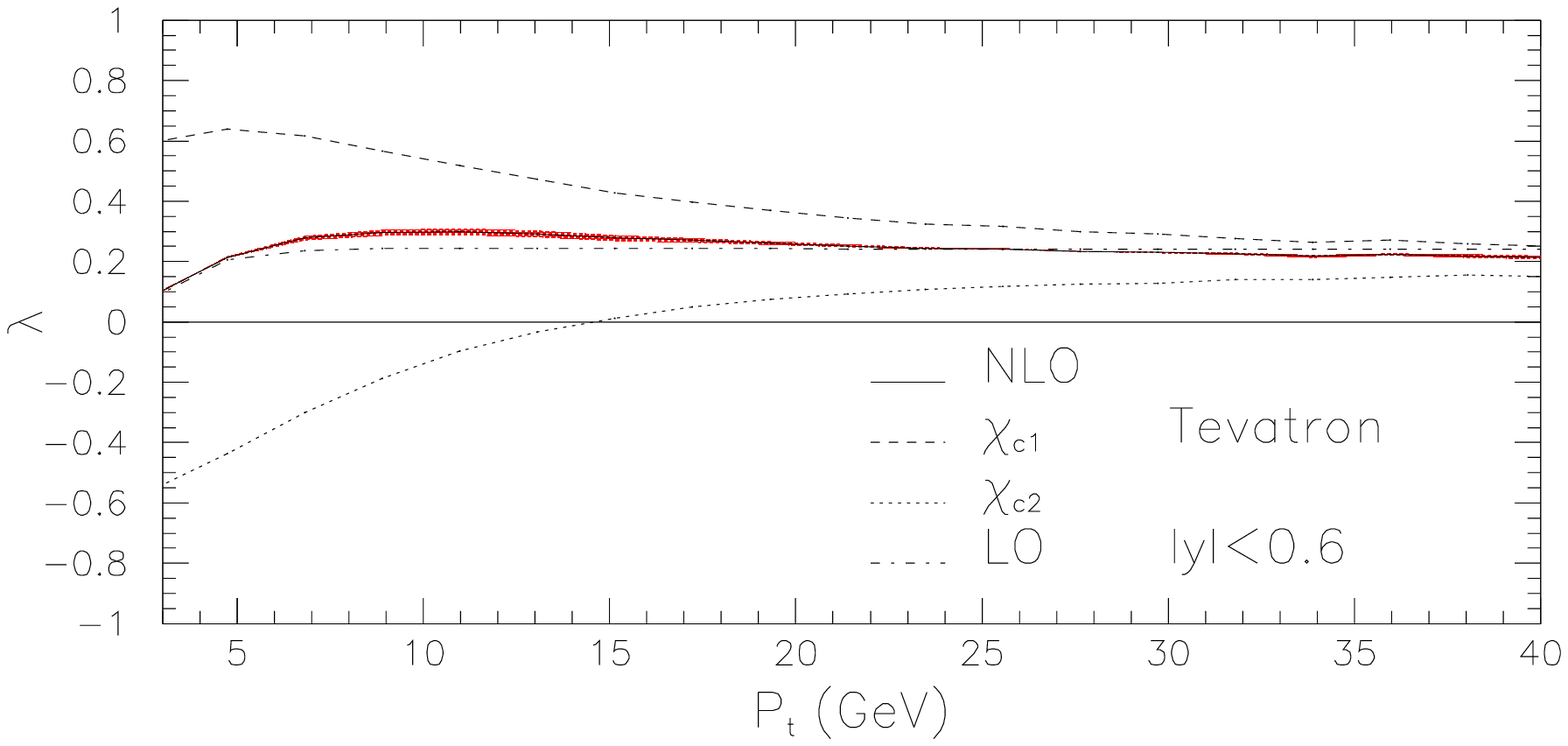}
\includegraphics*[scale=0.31]{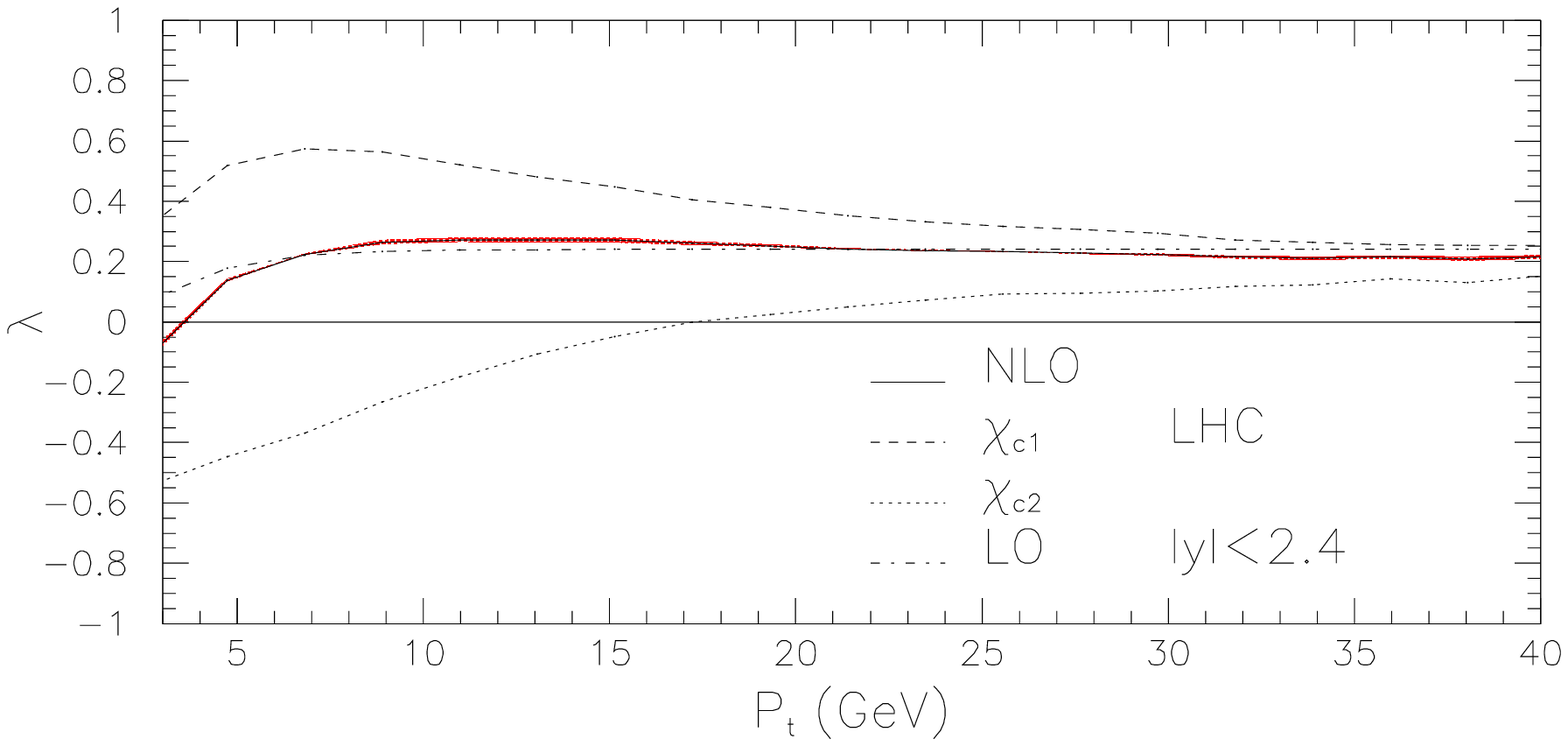}
\includegraphics*[scale=0.31]{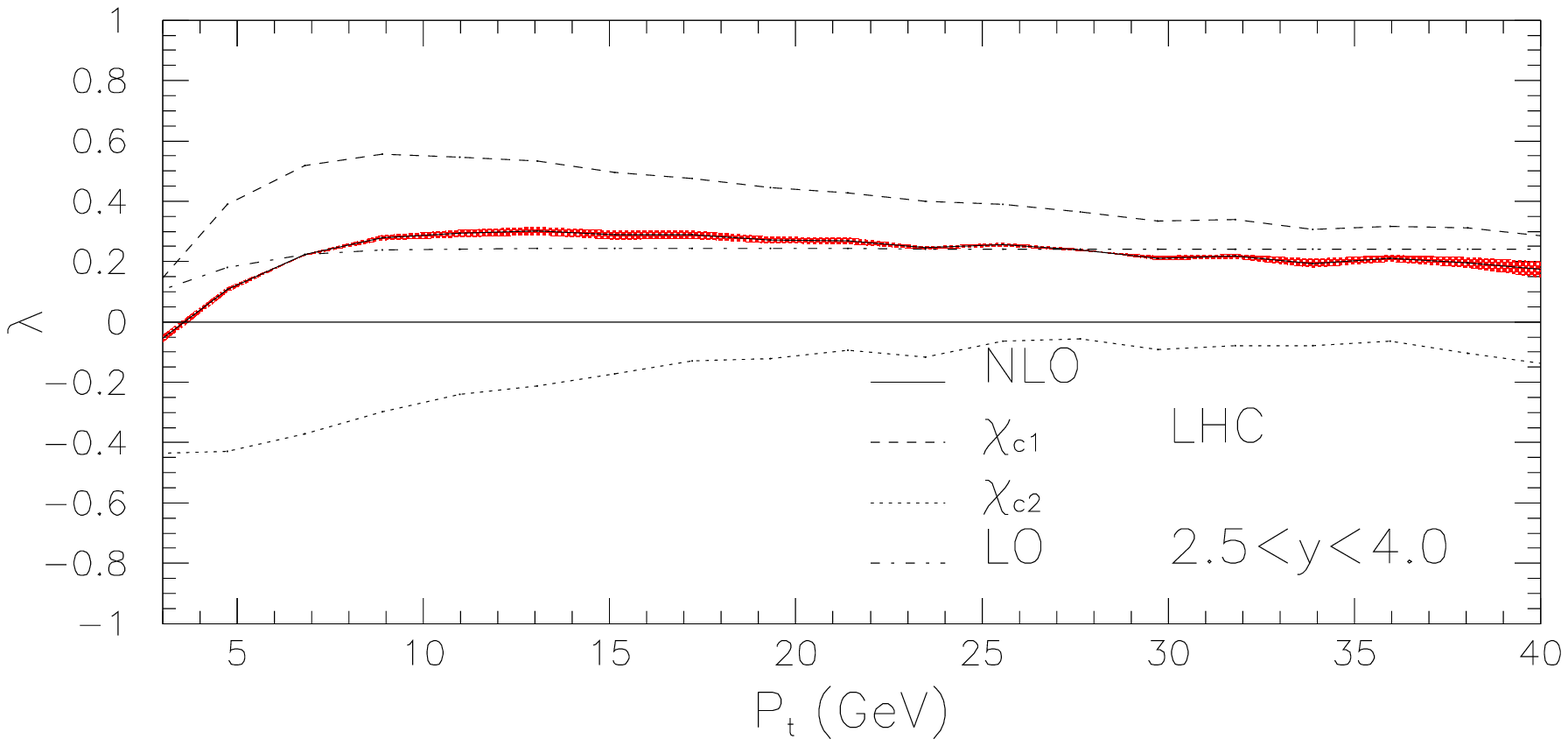}
\\
\includegraphics*[scale=0.31]{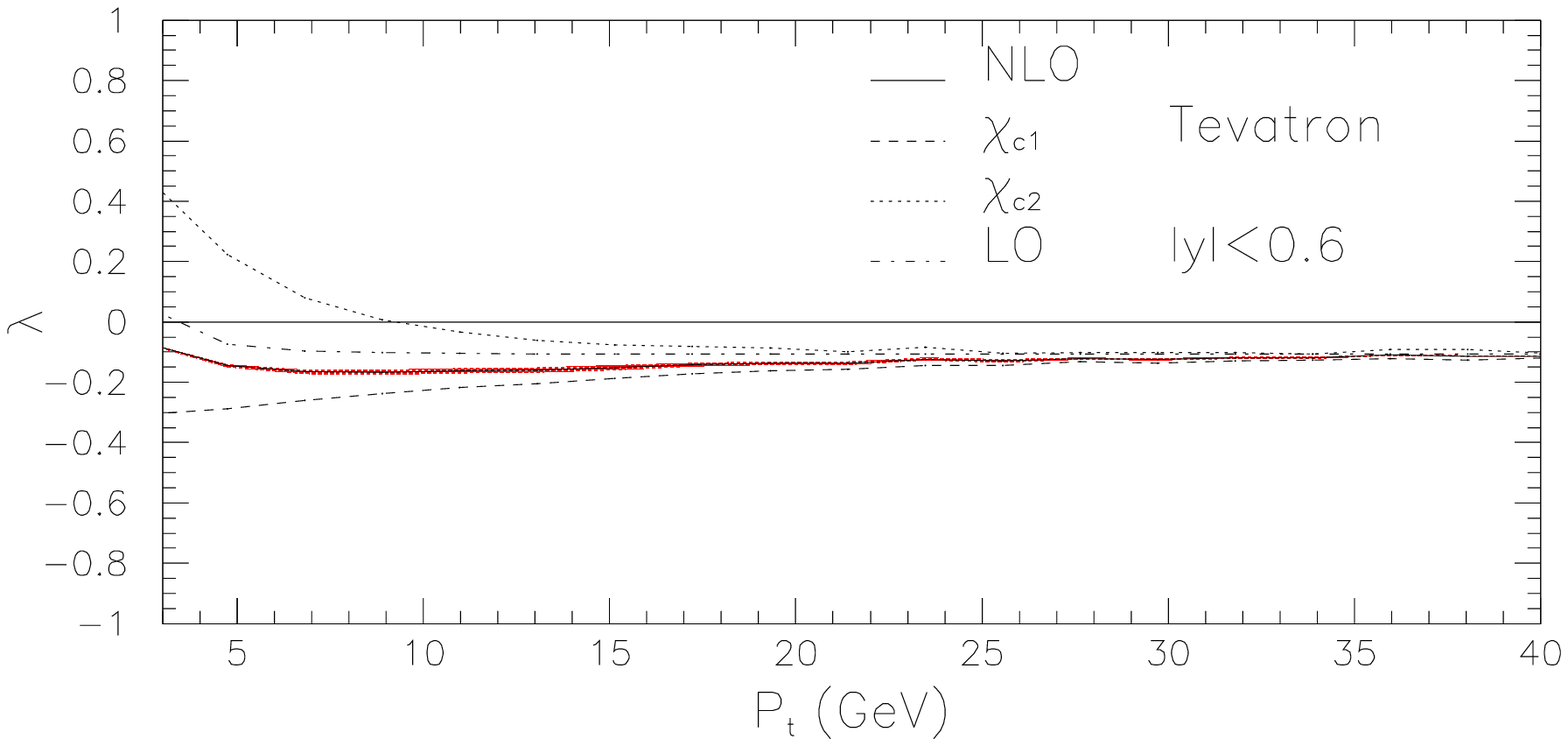}
\includegraphics*[scale=0.31]{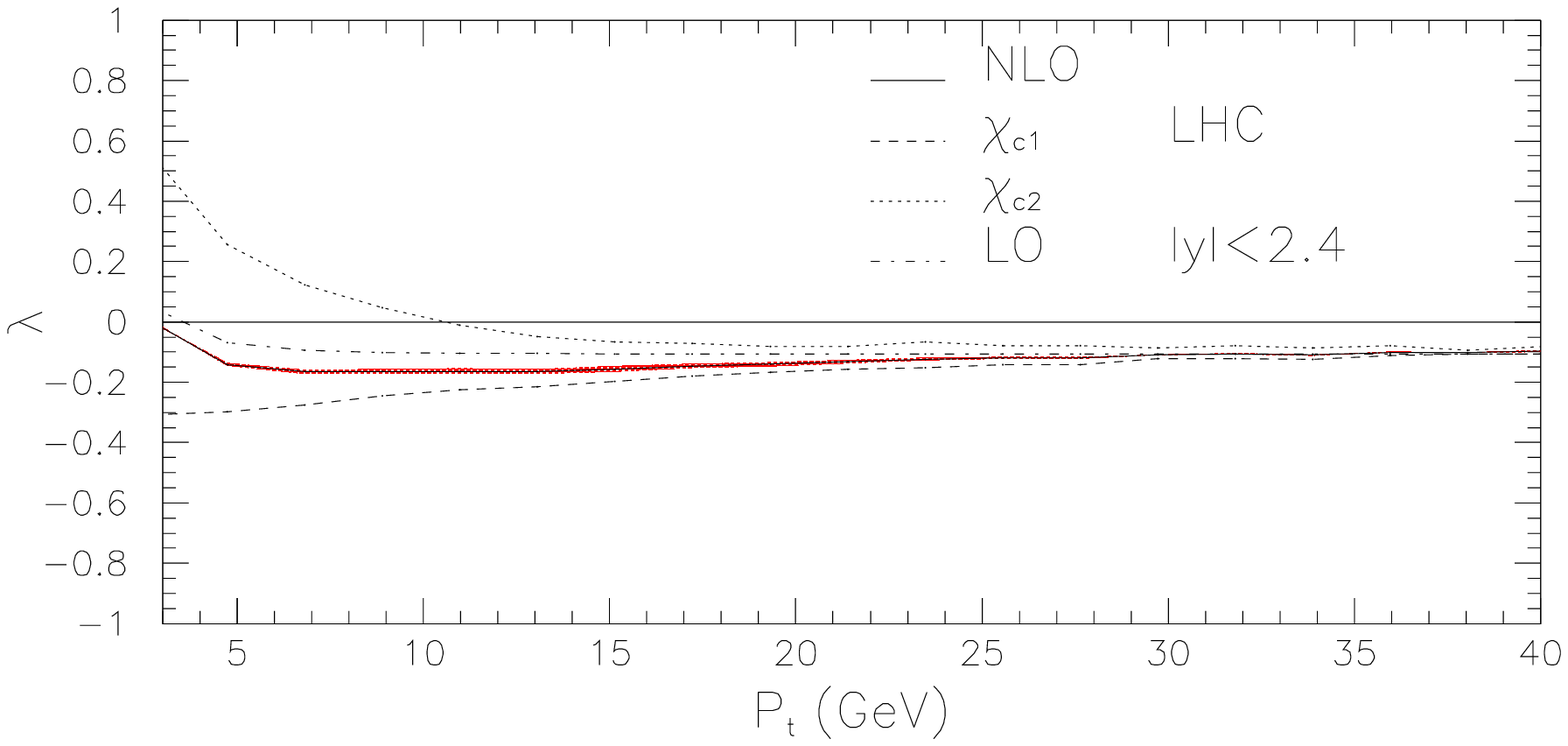}
\includegraphics*[scale=0.31]{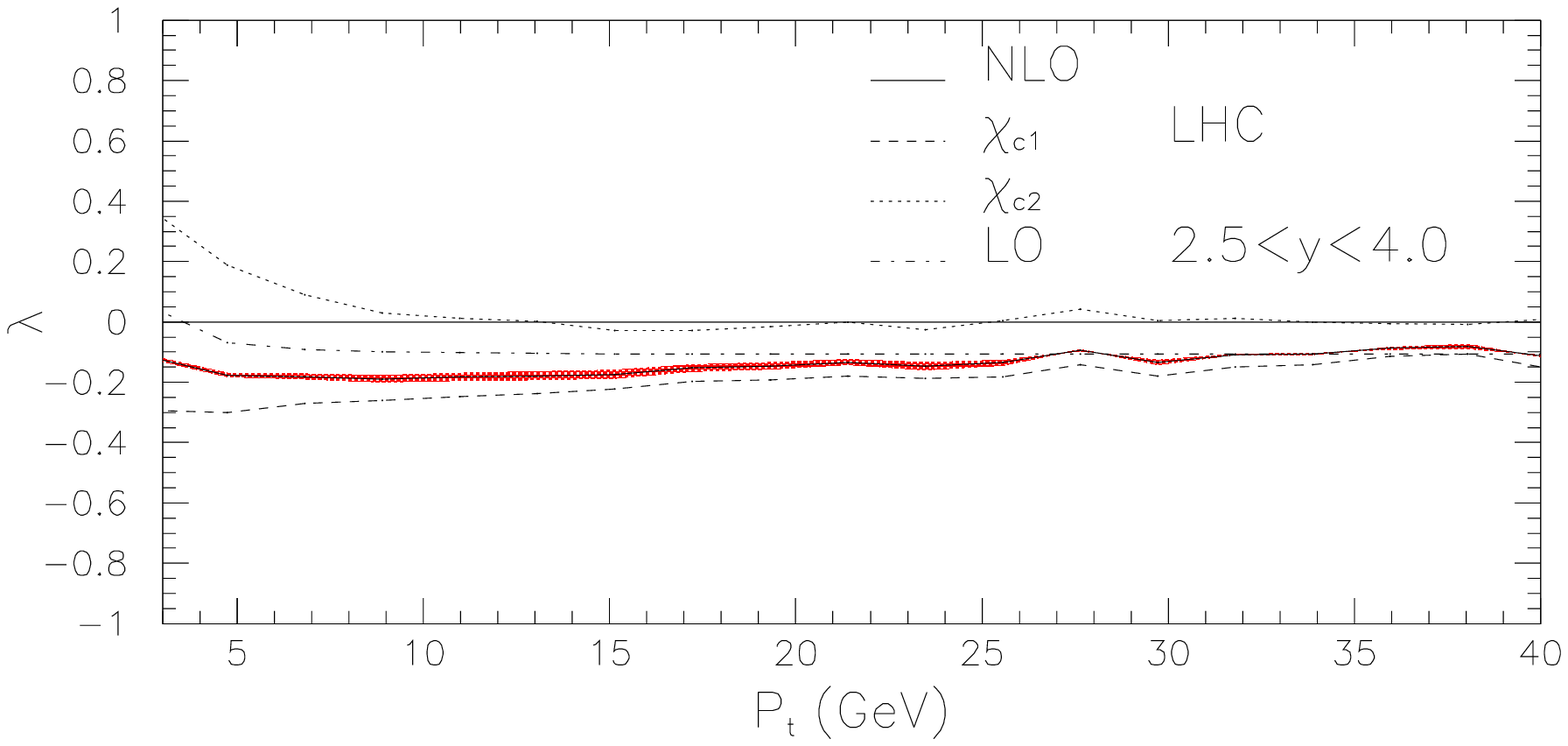}
\caption {\label{fig:polc}Polarization parameter $\lambda$ of $\jpsi$ from $\chi_{cJ}$ feeddown in helicity(up) and CS(down) frames.
}}
\end{figure*}

For $\chi_{cJ}$ production, 
the CO LDME is obtained as $\ME{\chi_{c0}}{}{(\OP{3}{S}{1}{8})} = (0.221\pm 0.012)\times 10^{-2}\gev^3$ 
with $\chi^2/\mathrm{d.o.f.}=2.57/10$ by fitting the CDF and LHCb data~\cite{Abe:1997yz, Aaij:2011jh, LHCb:2012af}.  

The CO LDMEs of $\psi(2s)$ are obtained from a combined fit of the CDF and LHCb experimental data~\cite{Aaltonen:2009dm} as
\bea
&(\ME{}{}{(\OP{1}{S}{0}{8})}, \ME{}{}{(\OP{3}{S}{1}{8})}, \frac{\ME{}{}{(\OP{3}{P}{0}{8})}}{m^2_c})
\equiv \dfrac{\co}{100} \gev^3 \label{eqn:psi_co},\\
&\co=(-0.012\pm 0.869, 0.34\pm 0.12, 0.42\pm 0.24) \NO
\eea
with $\chi^2/\mathrm{d.o.f.}=17.8/17$. 
The large uncertainty arises from approximately linear dependence of three short-distance parts. 
To clarify the situation,
a rotation matrix $V$ as discussed in Ref.~\cite{Ma:2010yw} is used to make fitting variables independent by introducing $\Lambda\equiv\co V$,
and $\Lambda$ is obtained with only independent error for each $\Lambda_i$ in the fit.
Then differential cross section $\md\sigma$ is obtained with
\bea
\md\sigma=\sum \co_i \md\hat{\sigma}_i =\sum \co V V^{-1}\md\hat{\sigma} =\sum \Lambda V^{-1} \md\hat{\sigma}. \NO \\ 
\Lambda=(0.16\pm 0.91, 0.47\pm 0.05, -0.21\pm 0.01), \label{eqn:psi2s_lambda} 
\label{eqn:rotation}
\eea
where $\Lambda$ is obtained with same $\chi^2/\mathrm{d.o.f.}$ but much smaller uncertainty.
With the values of $\Lambda$ and relation in Eq.~(\ref{eqn:rotation}), theoretical predictions with proper uncertainty can be made.

After the treatment of feeddown, the CO LDMEs for $\jpsi$ are obtained from a combined fit of the CDF and LHCb experimental data~\cite{Acosta:2004yw,Aaij:2011jh}.  
By using the same definition in Eq.~(\ref{eqn:psi_co}) for $\jpsi$, the fit gives
\bea
\co&=&(9.7\pm 0.9, -0.46\pm 0.13, -0.95\pm 0.25), \NO \\
\Lambda&=&(-9.6\pm 1.0, 1.7\pm 0.1, -0.37\pm 0.01), 
\eea
with $\chi^2/\mathrm{d.o.f.}=5.32/10$.
Thereafter, we use the error of each independent variable in $\Lambda$ to generate all the uncertainty bands in theoretical predictions.
It is clearly shown in the following figures that the uncertainty band is not too wide even with large uncertainty in $\Lambda$ in Eq.~(\ref{eqn:psi2s_lambda}). 

\begin{figure*}
\center{
\includegraphics*[scale=0.31]{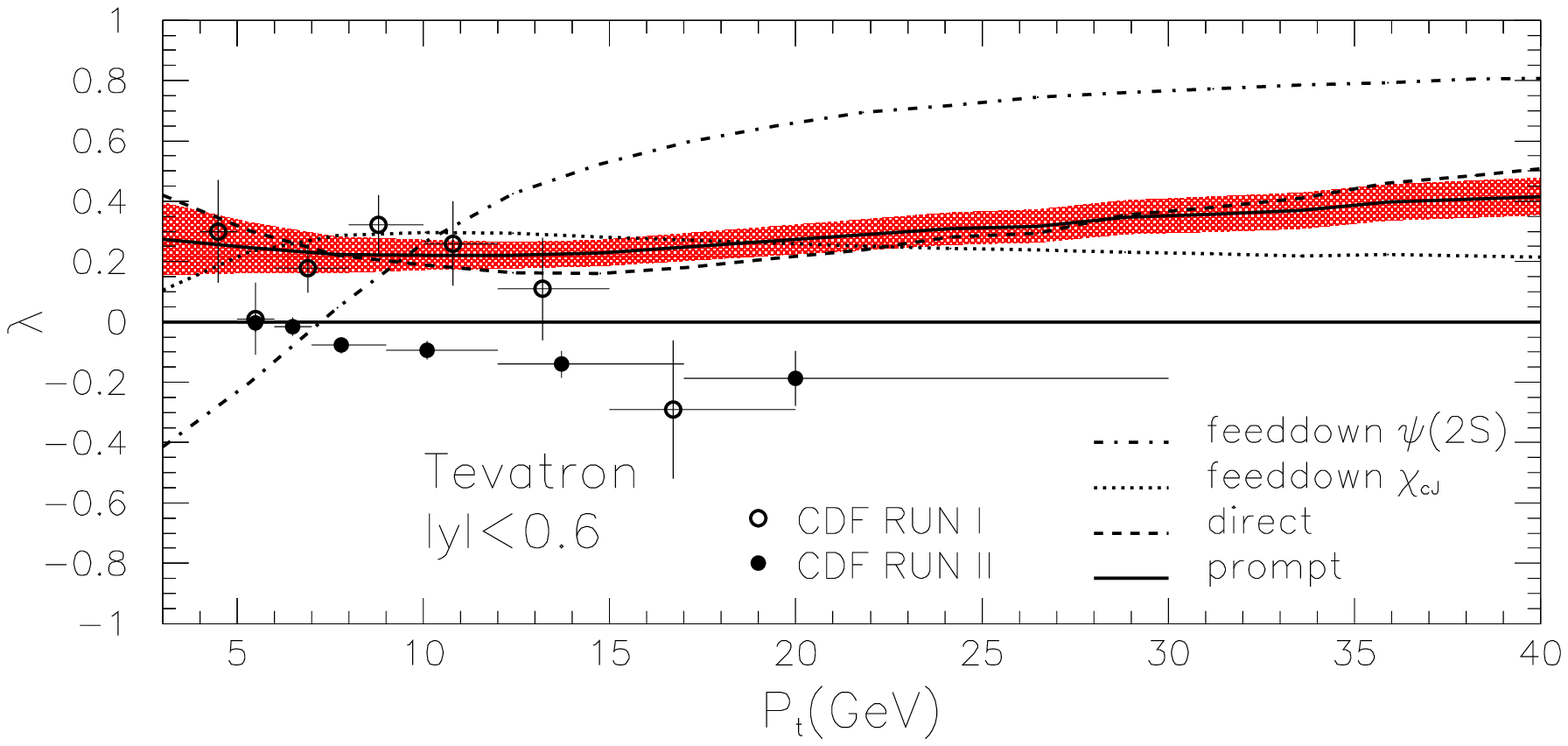}
\includegraphics*[scale=0.31]{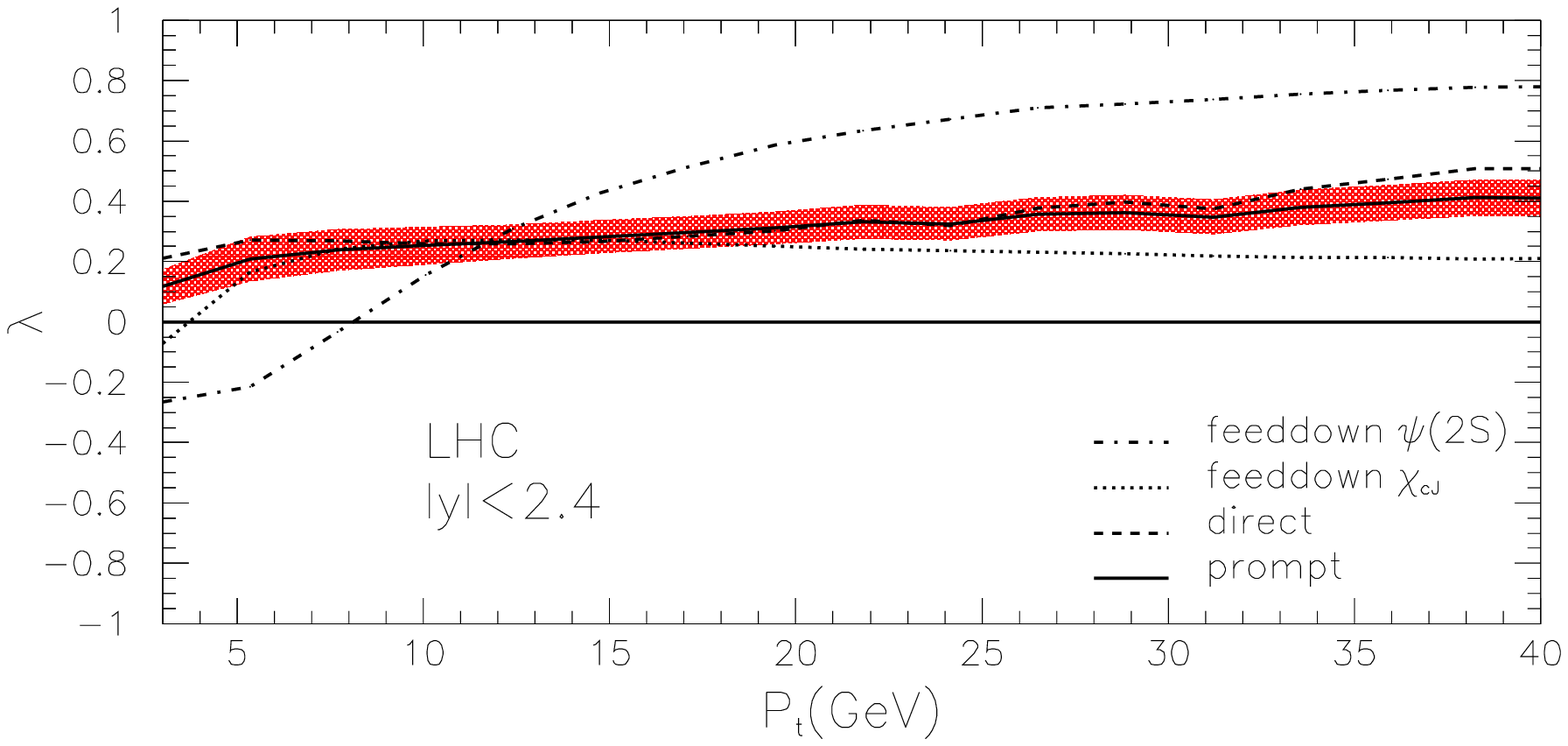}
\includegraphics*[scale=0.31]{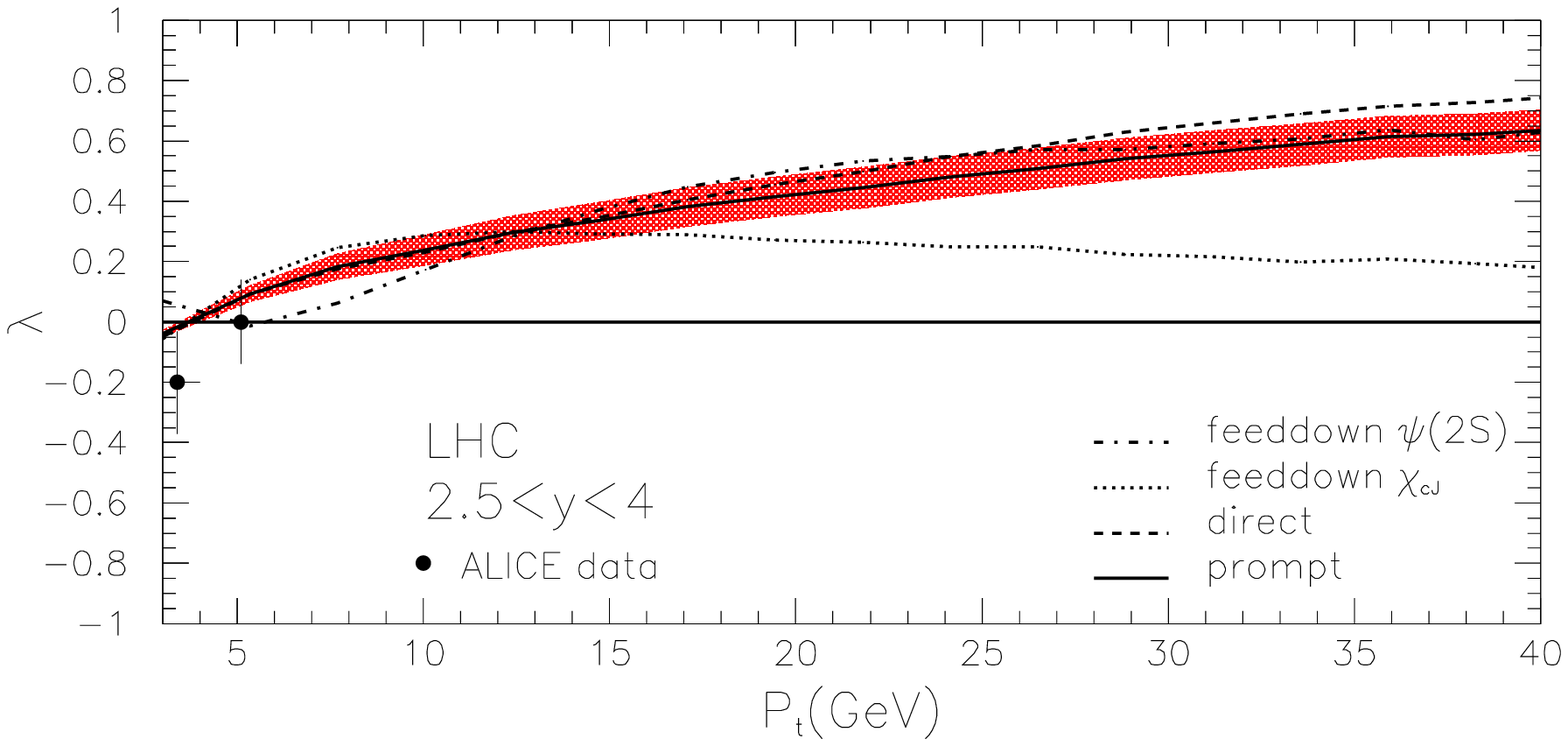}
\\
\includegraphics*[scale=0.31]{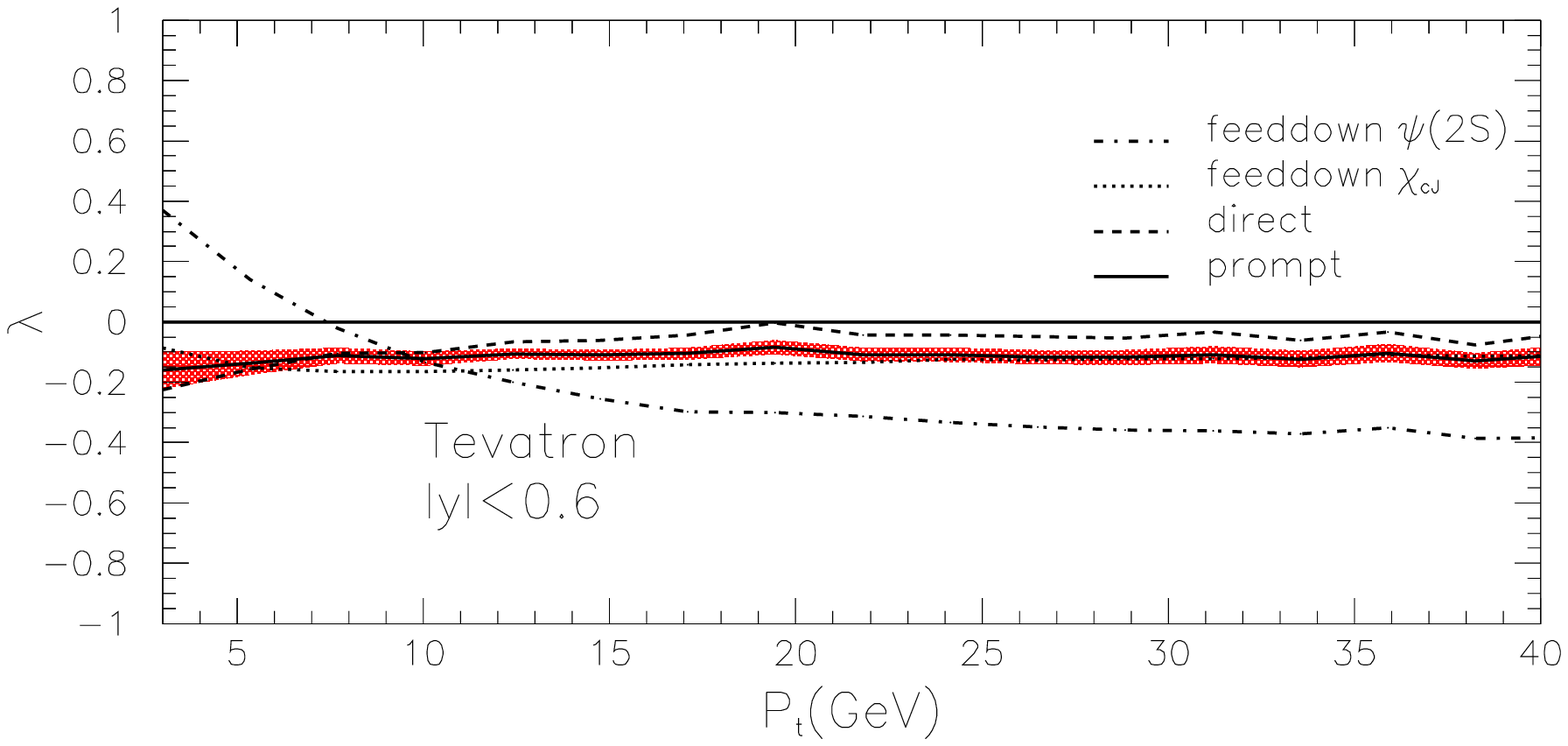}
\includegraphics*[scale=0.31]{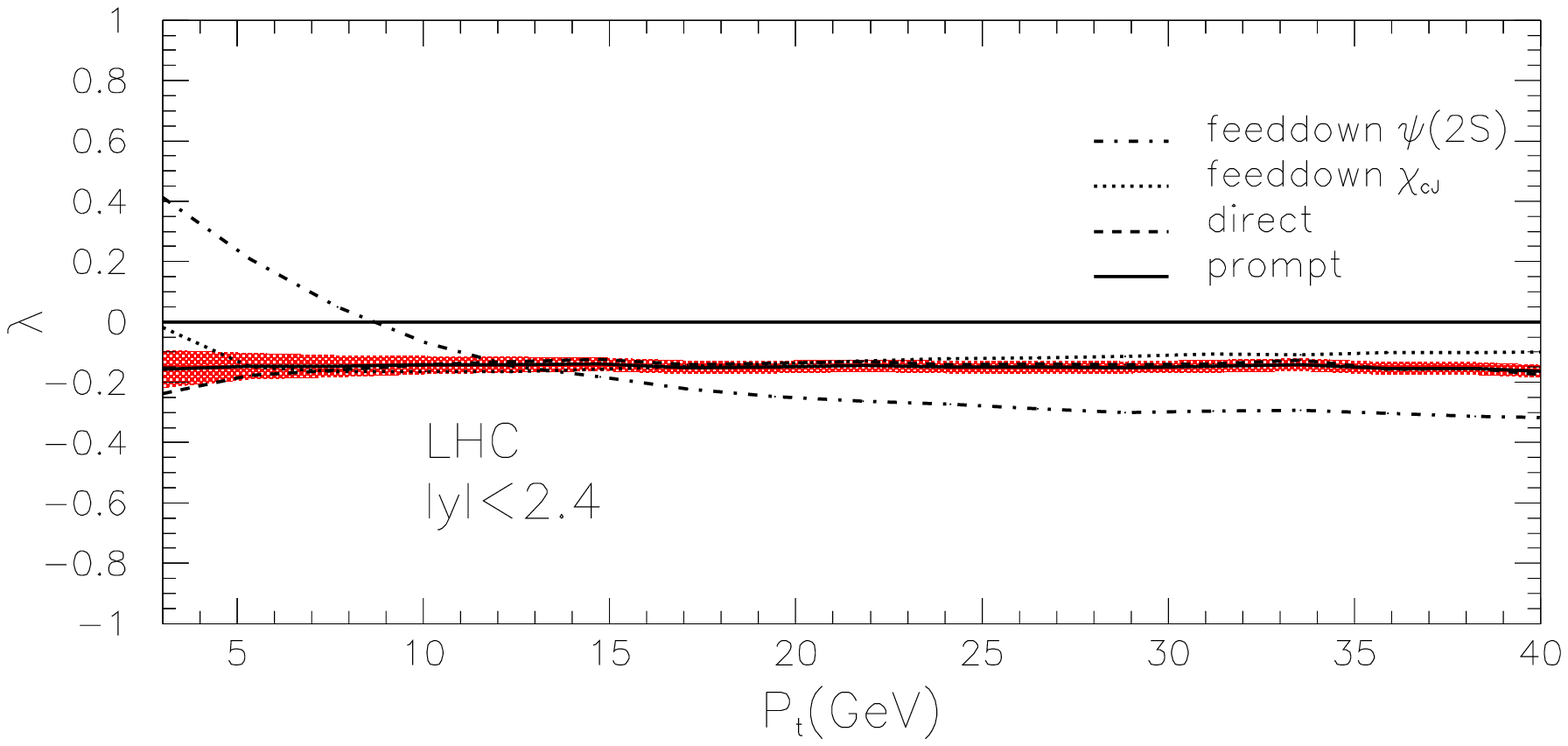}
\includegraphics*[scale=0.31]{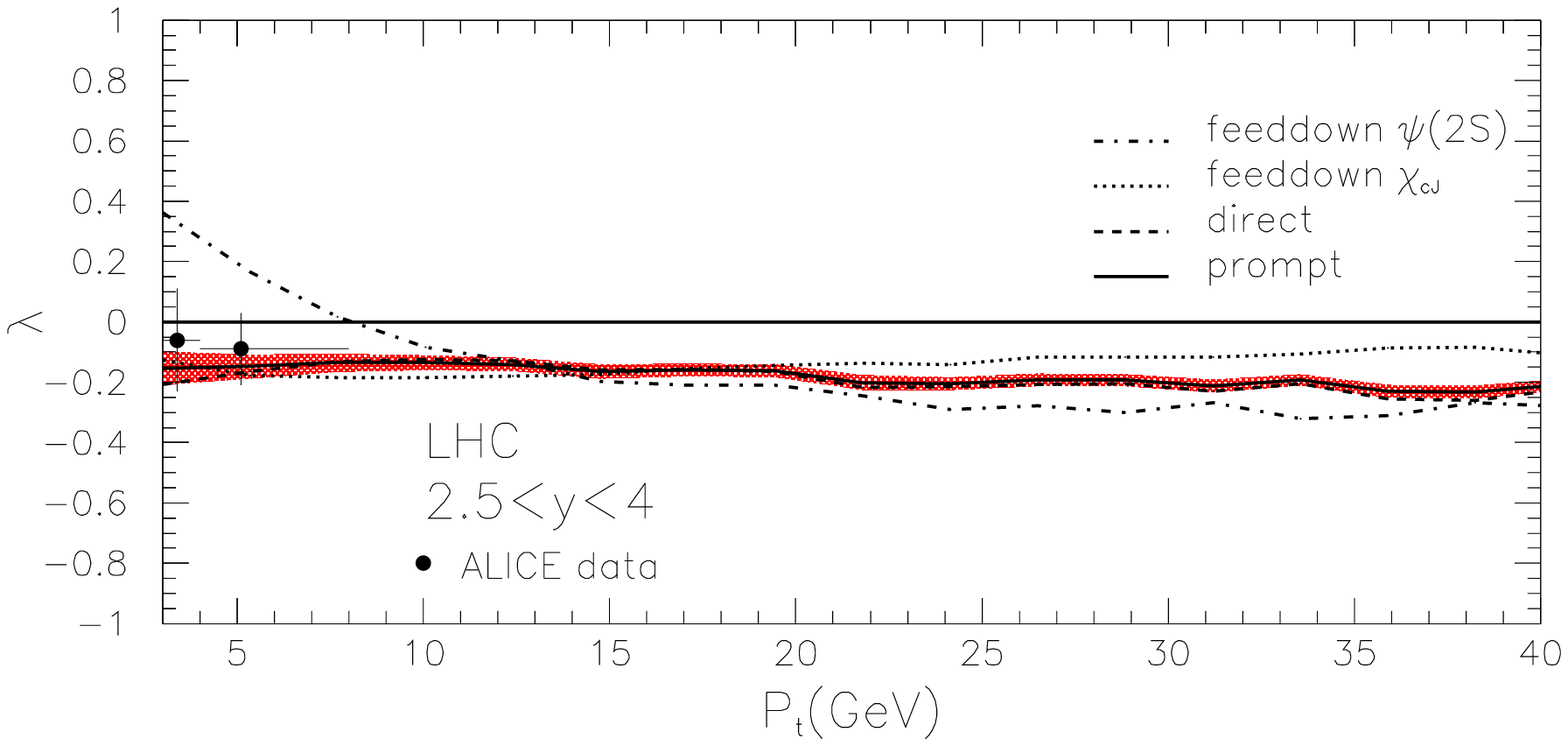}
\caption {\label{fig:pol}Polarization parameter $\lambda$ of prompt $\jpsi$ hadroproduction in helicity(up) and CS(down) frames.
The CDF and ALICE data are from Refs.~\cite{Affolder:2000nn,Abulencia:2007us}.
}}
\end{figure*}

In Fig.~\ref{fig:fit}, we find that the feeddown part contributes almost the same as the direct part in prompt $\jpsi$ yield when $p_t>25\gev$ at the Tevatron,
so as for $|y|<2.4$ when $p_t>30\gev$ at the LHC, but is less important in the forward range $4.5>y>2$.
The $\psi(2s)$ polarizations, shown in Fig.~\ref{fig:pol2}, 
go from longitudinal to transverse as $p_t$ increases in the helicity frame,
which has totally opposite trend with current CDF measurement~\cite{Abulencia:2007us},
and go from transverse to slight longitudinal in CS frame.
As is shown in Fig.~\ref{fig:polc},
the polarizations of $\jpsi$ from $\chi_{cJ}$ feeddown result in small transverse polarization ($\sim0.2$) in the helicity frame,
and slight longitudinal polarization ($\sim-0.1$) in the CS frame at large $p_t$.
Finally our theoretical predictions for prompt $\jpsi$ polarization in both helicity and CS frames are shown in Fig.~\ref{fig:pol}, 
in comparison with current existing measurements from the CDF and ALICE Collaborations.
In the forward rapidity region, our predictions are close to the ALICE measurement for inclusive $\jpsi$ production in both helicity and CS frames. 
In the central rapidity region, our results are in agreement with the CDF run I data (except two points), but in conflict with the CDF run II data.
However, we still cannot draw a definite conclusion since there is no way to judge these two measurements. 
From this point of view, this is another reason for us to exclude CDF data on $J/\psi$ polarization in our fitting.
With the optimized analysis method to measure more $\jpsi$ polarization information with two different frames (helicity and CS) as used in the ALICE measurement,
it is expected that $\jpsi$ and $\psi(2s)$ polarization measurements at the LHC would help to solve the polarization puzzle or clarify the situation.  

The polarization predictions in Ref.~\cite{Butenschoen:2012px} and \cite{Chao:2012iv} are for direct $J/\psi$ production. 
Comparing them with the CDF (ALICE) measurements for prompt (inclusive) $\jpsi$ production and making a definite conclusion
can be considered as reasonable when and only when the feeddown contribution is negligible. 
Our calculation shows that the feeddown contribution is very important and cannot be neglected.
Our predictions are in better agreement with the ALICE measurement than theirs in Ref~\cite{Butenschoen:2012px}.
By excluding the feeddown contribution,
our polarization predictions for direct $J/\psi$ production are consistent with the results in Ref.~\cite{Butenschoen:2012px} and \cite{Chao:2012iv} by using their fitted LDMEs.

In summary,
we presented the first complete NLO study on polarization for prompt $\jpsi$ and $\psi(2s)$ hadroproduction in both the helicity and CS frames
by including feeddown contributions from higher charmonia $\chi_c$ and $\psi(2s)$ based on the NRQCD factorization scheme.  
It is found that there is no physical solution for a combined fit of both $\jpsi$ yield and polarization measurements by CDF at the Tevatron (run II) for prompt $\jpsi$ hadroproduction,
while the previous fit~\cite{Chao:2012iv} without considering the polarization from feeddown of $\chi_c$ and $\psi(2s)$ can be reproduced.
It means that the polarization contributed from $\chi_c$ and $\psi(2s)$ feeddown is a very important part and must be included to solve or clarify the $\jpsi$ polarization puzzle. 
Therefore, we choose to use the new CO LDMEs from a combined fit of $\jpsi$ yield measurements at the Tevatron and LHC with $p_t>7\gev$.
And the NLO theoretical predictions on polarization for prompt $\jpsi$ are presented at the Tevatron and LHC.
The results are in agreement with the CDF run I data (except two points), but in conflict with the CDF run II data.
It is close to the ALICE measurement at the LHC although the measurement is for inclusive $\jpsi$.
Finally, it is clear that the polarization measurements at the LHC are very important to clarify the long-standing $\jpsi$ polarization puzzle. 

We are thankful for help from the Deepcomp7000 project of the Supercomputing Center, CNIC, CAS.
This work is supported by the National Natural Science Foundation of China (No.~10979056, No. 10935012, and No. 11005137), and by CAS under Project No. INFO-115-B01.

\end{document}